\documentclass[a4paper,12pt]{article}

\usepackage{geometry}
\usepackage{graphicx}
\usepackage{enumitem}
\usepackage{xcolor}
\usepackage{soul} 
\usepackage{amsmath,amssymb}
\usepackage{tikz}
\usepackage{mathtools}
\usepackage{siunitx}
\usepackage{tocloft}
\usepackage{titlesec}
\usepackage{wasysym}
\usepackage{hyperref}
\usepackage{multirow}
\usepackage{ragged2e}
\usepackage{booktabs}
\usepackage{array}
\usepackage{pifont}
\usepackage{makecell}
\usepackage{threeparttable}
\usepackage[square,numbers]{natbib}
\usepackage[hang,flushmargin]{footmisc}
\usepackage{tabularx}
\usepackage{setspace}

\sethlcolor{green!30}
\date{}

\begin{document}
	
	\onehalfspacing
	\begin{titlepage}
		\thispagestyle{empty}
		
		\vspace{.4cm}
		\begin{center}
			{\Large \textbf{
					Impact of Higgs precision measurements at the LHC and  FCC-ee on the spectrum of composite Higgs models
			}} 
		\end{center}
		
		\vspace*{15mm}
		
		\begin{center}
			{Reza Asgharzadeh Jelodar$^a$, Kazem Bitaghsir Fadafan$^a$ and Giacomo Cacciapaglia$^b$}
		\end{center}
		\vspace*{1cm}
		
		\begin{center}
			{\it $^a$ Faculty of Physics, Shahrood University of Technology, P.O.Box 3619995161, Shahrood, Iran\\
				$^b$ Laboratoire de Physique Th\'eorique et Hautes \'Energies (LPTHE), UMR 7589, Sorbonne Universit\'e \& CNRS, 4 place Jussieu, 75252 Paris Cedex 05, France\\
			}
			
			\vspace*{0.5cm}
			{E-mails: {\tt rezaasgharzadehjelodar@shahroodut.ac.ir, bitaghsir@shahroodut.ac.ir, cacciapa@lpthe.jussieu.fr}}
		\end{center} 
		
\begin{abstract}
	We investigate the minimal composite Higgs model based on the symmetry-breaking pattern \(\mathrm{SU}(4)\rightarrow\mathrm{Sp}(4)\), where electroweak symmetry breaking is governed by the vacuum alignment angle \(\theta\). Through the leading-order relations \(\kappa_V=\cos\theta\) and \(m_\eta=m_h/\sin\theta\), precision measurements of Higgs couplings are translated into direct constraints on the vacuum structure and the singlet pseudo-Nambu--Goldstone boson mass. Using the current ATLAS Run-2 measurement of \(\kappa_V\), we construct a Bayesian posterior for \(\theta\), including the Jacobian associated with the transformation from \(\kappa_V\) to \(\theta\), and validate the results through an independent frequentist \(\Delta\chi^2\) analysis. The same framework is then applied to the projected sensitivities of the High-Luminosity LHC and FCC-ee. The current data imply a conservative lower bound of approximately \(440~\mathrm{GeV}\) on the singlet mass at the \(95\%\) credibility level, while the projected sensitivities improve this limit to about \(600~\mathrm{GeV}\) at the HL-LHC and beyond \(2~\mathrm{TeV}\) at FCC-ee. The close agreement between the Bayesian and frequentist determinations demonstrates the robustness of the extracted constraints. These results show that future Higgs precision measurements will probe the vacuum alignment of the minimal \(\mathrm{SU}(4)/\mathrm{Sp}(4)\) composite Higgs model with unprecedented sensitivity, placing increasingly stringent constraints on the allowed parameter space and establishing the singlet scalar as a compelling target for upcoming collider programs.
\end{abstract}
	\end{titlepage}
	\newpage
	
	\tableofcontents
	\noindent
	\hrulefill	
	
	\section{Introduction}
	\label{sec:introduction}
	
	The discovery of the Higgs boson with a mass of approximately \(125\ \mathrm{GeV}\) at the Large Hadron Collider (LHC) marked a milestone in our understanding of the electroweak symmetry breaking mechanism \cite{ATLAS2012, CMS2012}. However, this discovery also sharpened a more fundamental question: why is the Higgs mass stable against large quantum corrections from much higher scales? In the Standard Model, the Higgs is a fundamental scalar whose mass receives quadratically divergent radiative corrections, requiring a delicate cancellation between the bare mass and loop contributions. This tension is commonly known as the hierarchy problem \cite{Weinberg1976, Susskind1979}.
	
	A compelling resolution is to abandon the idea that the Higgs is elementary and instead interpret it as a composite state, specifically as a pseudo-Nambu--Goldstone boson (pNGB) arising from new strong dynamics at a higher scale \cite{KaplanGeorgi1984, KaplanGeorgiDimopoulos1984}. In such composite Higgs models, the Higgs mass is naturally protected by the approximate global symmetry of the strong sector, while the observed Higgs emerges as a low-energy effective degree of freedom \cite{Contino2010, Panico:2015jxa}. Among the many realizations, a minimal model based on the symmetry-breaking pattern \(\mathrm{SU}(4) \to \mathrm{Sp}(4)\) can be realized from a fundamental \(\mathrm{SU}(2)\) gauge theory with two Dirac fermions \cite{Cacciapaglia2014, Hietanen2014}. The coset structure yields five pNGBs, which decompose under the electroweak subgroup into a Higgs doublet and a neutral singlet scalar \(\eta\). The phenomenology is controlled by a single vacuum alignment angle \(\theta\), which determines both the Higgs coupling modifications and the singlet mass through the leading-order relations
	
	\begin{equation}
		\kappa_V \equiv \frac{g_{hVV}}{g_{hVV}^{\mathrm{SM}}} = \cos\theta, \qquad
		m_\eta = \frac{m_h}{\sin\theta},
	\end{equation}
	
	where \(\kappa_V\) is the Higgs coupling modifier to vector bosons \cite{Cacciapaglia2014, Contino2010}. A small \(\theta\), corresponding to near-perfect alignment, implies a heavy singlet and Higgs couplings close to Standard Model values. And also the electroweak scale is set by \( v = 2\sqrt{2} f \sin\theta \) (we adopt this convention throughout; an alternative \( v = f'\sin\theta \) exists in the literature). We recall that the form of $\kappa_V$ is universal to all model of this kind~\cite{1805.00489, 1809.09126}, while the $\eta$ mass is model-dependent.
	
	Our purpose is to translate precision measurements of \(\kappa_V\) -- both current and projected -- into quantitative constraints on \(\theta\) and, through the relation above, on the singlet mass \(m_\eta\). We take as our starting point the ATLAS Run-2 combined measurement \(\kappa_V = 1.035 \pm 0.031\) \cite{ATLAS:2022vkf}, and extend the analysis to the projected sensitivities of the High-Luminosity LHC \cite{ATLAS:2025hllhc} and FCC-ee \cite{FCC:2025feasibility}. For each scenario, we construct a Bayesian posterior for \(\theta\), including the Jacobian factor arising from the nonlinear transformation \(\kappa_V \to \theta\), and cross-check the results against an independent frequentist \(\Delta\chi^2\) profile.
	
	The Run-2 bound represents an interesting limiting case: since the central value \(\mu = 1.035\) lies above the physical boundary \(\kappa_V \leq 1\), the posterior is dominated by the truncation at \(\kappa_V = 1\) rather than by the data peak. This yields conservative limits, \(\theta_{95\%} \simeq 0.291\) and \(m_\eta \gtrsim 440\ \mathrm{GeV}\) from the Bayesian analysis, with the frequentist cross-check giving \(\theta \simeq 0.232\) and \(m_\eta \gtrsim 540\ \mathrm{GeV}\). At HL-LHC, the bound tightens to \(\theta_{95\%} \simeq 0.21\), pushing \(m_\eta\) above \(600\ \mathrm{GeV}\). The improvement at FCC-ee is far more dramatic, with \(\theta_{95\%} \simeq 0.06\) and \(m_\eta\) excluded below approximately \(2\ \mathrm{TeV}\). The close agreement between Bayesian and frequentist limits confirms the robustness of these conclusions.

	Current Higgs coupling measurements already disfavour large vacuum misalignment, corresponding to a relatively light \(\eta\), while the projected FCC-ee precision would exclude nearly the entire remaining parameter space with sizable misalignment. Consequently, the singlet scalar emerges as a well-defined target for future Higgs precision programs. Our results highlight the impact of high-precision Higgs measurements in constraining the spectrum of
	composite Higgs model.  This analysis could be eaily extended to other models and additional resonances, which 	are a natural target for the complete LHC programme and a 100 TeV hadronic collider project.

	The paper is organized as follows. Section~\ref{sec:theoretical_framework} reviews the \(\mathrm{SU}(4)/\mathrm{Sp}(4)\) composite Higgs model and derives the key relations between \(\theta\), \(\kappa_V\), and \(m_\eta\). Section~\ref{sec:numerical_analysis} presents the statistical methodology and the extraction of constraints from current and projected data. Section~\ref{sec:phenomenology} discusses the phenomenological implications and compares the reach of Run-2, HL-LHC, and FCC-ee. Section~\ref{sec:conclusion} summarizes our findings and outlines future directions.
	
	\section{Composite Higgs and Vacuum Alignment}
	\label{sec:theoretical_framework}
	
	As discussed above, the central idea of composite Higgs models is that the Higgs boson is not an elementary scalar but a pseudo-Nambu--Goldstone boson (pNGB) arising from the spontaneous breaking of a global symmetry in a new strongly coupled sector \cite{KaplanGeorgi1984, KaplanGeorgiDimopoulos1984, Contino2010, Panico:2015jxa}. This structure naturally protects the Higgs mass from large quadratic corrections, addressing the hierarchy problem and, in part, the little hierarchy \cite{hep-ph/0405040}.
	
	Among the many possible realizations, the minimal composite Higgs model based on the symmetry-breaking pattern \(\mathrm{SU}(4) \to \mathrm{Sp}(4)\) stands out for its simplicity and phenomenological richness \cite{Cacciapaglia2014, Galloway2010, Gripaios2009}. This pattern can be realized from a fundamental \(\mathrm{SU}(2)\) gauge theory with two Dirac fermions in the fundamental representation \cite{Hietanen2014, Lewis:2011zb,Belyaev:2019ybr,BitaghsirFadafan:2018efw}. The breaking \(\mathrm{SU}(4) \to \mathrm{Sp}(4)\) produces five pNGBs, which decompose under the electroweak gauge group \(\mathrm{SU}(2)_L \times \mathrm{U}(1)_Y\) as a Higgs doublet and a neutral singlet scalar \(\eta\). The explicit derivation of this symmetry-breaking pattern from the underlying $\mathrm{SU}(2)$ gauge theory with two Dirac fermions, including the construction of the $\mathrm{SU}(4)$-covariant fermion bilinear and the counting of the ten unbroken generators of $\mathrm{Sp}(4)$, is given in  Appendix~\ref{app:lagrangian_rewrite}.
	
	The vacuum of the strong sector is characterized by a \(4 \times 4\) antisymmetric matrix \(\Sigma_0\) that transforms under \(\mathrm{SU}(4)\) as \(\Sigma_0 \to U \Sigma_0 U^T\). Two distinct vacuum configurations are relevant for electroweak symmetry breaking \cite{Galloway2010}: the electroweak-preserving vacuum \(\Sigma_B\) and the electroweak-breaking vacuum \(\Sigma_H\). The physical vacuum can be characterized as a linear combination of these two configurations:,
	
	\begin{equation}
		\Sigma_0 = \cos\theta \, \Sigma_B + \sin\theta \, \Sigma_H,
		\label{eq:vacuum_alignment}
	\end{equation}
	
	where \(\theta\) is the vacuum alignment angle. This angle interpolates between the unbroken phase (\(\theta = 0\)) and the  Technicolor-like limit (\(\theta = \pi/2\)). The composite Higgs regime corresponds to small \(\theta \ll 1\). The kinetic term for the pNGBs, minimally coupled to the electroweak gauge bosons, yields the masses of the \(W\) and \(Z\) bosons \cite{Contino2010}:
	
	\begin{equation}
		m_W^2 = 2g^2 f^2 \sin^2\theta, \qquad m_Z^2 = 2(g^2 + g'^2) f^2 \sin^2\theta,
	\end{equation}
	
	where \(f\) is the pNGB decay constant. Comparing with the Standard Model relation \(m_W = gv/2\) gives the electroweak vacuum expectation value
	
	\begin{equation}
		v = 2\sqrt{2} f \sin\theta.
		\label{eq:v_theta}
	\end{equation}
	Three pNGBs remain exact Goldstones and provide the lingitudinal polarizations of $W^\pm$ and $Z$.
	The physical Higgs boson \(h\) emerges as the fluctuation of the field corresponding to the broken generator , while \(\eta\) corresponds to the orthogonal broken generator . Expanding the kinetic term to linear order in the fields gives the Higgs couplings to the \(W\) and \(Z\) bosons \cite{Cacciapaglia2014}:
	
	\begin{equation}
		g_{hWW} = g_{hWW}^{\mathrm{SM}} \cos\theta, \qquad g_{hZZ} = g_{hZZ}^{\mathrm{SM}} \cos\theta.
	\end{equation}
	
	Thus, the Higgs coupling modifier is directly tied to the vacuum alignment angle:
	
	\begin{equation}
		\kappa_V \equiv \frac{g_{hVV}}{g_{hVV}^{\mathrm{SM}}} = \cos\theta.
		\label{eq:kappa_theta}
	\end{equation}
	
	This relation is exact at leading order in the effective theory and provides the crucial link between experimental measurements and the model parameters. Since \(\kappa_V \leq 1\) by construction, any deviation from unity signals vacuum misalignment.
	
	While the Higgs couplings follow from the kinetic term, the physical masses of the pNGBs are determined by the effective potential generated by explicit breaking of the global \(\mathrm{SU}(4)\) symmetry. The dominant sources of this breaking are the electroweak gauge interactions and, most importantly, the top-quark Yukawa coupling \cite{Cacciapaglia2014, Barnard:2013zea}.
	
	The effective potential can be written as the sum of gauge, top-loop, and explicit mass contributions. Following Ref.~\cite{Cacciapaglia2014}, the potential simplifies to
	
	\begin{equation}
		V(\theta) = X_t \cos^2\theta - 4X_m \cos\theta + \text{const.},
	\end{equation}
	
	where \(X_t\) encodes the top-loop and gauge contributions and \(X_m\) encodes the explicit breaking. Minimizing with respect to \(\theta\) gives the equilibrium condition
	
	\begin{equation}
		\cos\theta_{\min} = \frac{2X_m}{X_t}.
		\label{eq:min_cond}
	\end{equation}
	
	The second derivatives of the potential at the minimum yield the mass eigenvalues:
	
	\begin{equation}
		m_h^2 = \frac{f^2}{4} X_t \sin^2\theta, \qquad m_\eta^2 = \frac{f^2}{4} X_t.
		\label{eq:masses}
	\end{equation}
	
	The crucial observation is that both masses share the same prefactor \(f^2 X_t/4\); the only difference is the geometric factor \(\sin^2\theta\) multiplying the Higgs mass. This reflects the distinct group-theoretic roles of the two fields: the Higgs direction couples to the explicit-breaking sources in a way that tracks the alignment angle, while the singlet direction is orthogonal and retains its mass even as \(\theta \to 0\).
	
	Dividing the two mass expressions yields a purely geometric relation:
	
	\begin{equation}
		\frac{m_h^2}{m_\eta^2} = \sin^2\theta.
	\end{equation}
	
	Taking the positive root and substituting the measured Higgs mass \(m_h \simeq 125\ \mathrm{GeV}\) gives
	
	\begin{equation}
		m_\eta = \frac{m_h}{\sin\theta}.
		\label{eq:meta_theta}
	\end{equation}
	This relation is central to the phenomenology of the model: a small vacuum alignment angle (corresponding to \(\kappa_V \simeq 1\)) implies a heavy singlet scalar.
	
	 The full derivation of this relation from the effective potential generated by explicit $\mathrm{SU}(4)$ breaking -- including the gauge, top-quark, and explicit mass contributions, and the resulting expressions for $m_h^2$ and $m_\eta^2$ -- is given in  Appendix~\ref{app:mass_relation_detail}. Model-dependent corrections to this prediction may arise from additional spurions breaking the global SU(4)	 symmetry, or from a different implementation of the top mass. Throughout this work we treat $\theta$ as a free phenomenological parameter constrained directly by Higgs coupling data, following the standard approach of Refs.~\cite{Cacciapaglia2014, Barnard:2013zea}, rather than deriving its value from the potential parameters $X_t$ and $X_m$; higher-order corrections to Eq.~\eqref{eq:meta_theta} are $\mathcal{O}(\theta^2)$ and are neglected here.

	The three relations derived above form the foundation of our analysis:
	
	\begin{align}
		\kappa_V &= \cos\theta, \label{eq:key1}\\
		v &= 2\sqrt{2} f \sin\theta, \label{eq:key2}\\
		m_\eta &= \frac{m_h}{\sin\theta}. \label{eq:key3}
	\end{align}
	
	Equation \eqref{eq:key1} connects the observable Higgs coupling modifier to the vacuum alignment angle. Equation \eqref{eq:key2} relates the electroweak scale to the strong dynamics scale \(f\). Equation \eqref{eq:key3} determines the mass of the singlet scalar in terms of \(\theta\). These relations have profound implications. Precision measurements of \(\kappa_V\) directly constrain \(\theta\). Since \(m_\eta = m_h / \sin\theta\), a small \(\theta\)—which is increasingly preferred by precise Higgs coupling measurements—implies a heavy singlet. Conversely, a light \(\eta\) would require significant vacuum misalignment (\(\theta \sim \mathcal{O}(1)\)), which would be accompanied by a visible reduction in the Higgs couplings to gauge bosons.
	
	This interplay between Higgs coupling measurements and the singlet mass provides the central motivation for the numerical analysis presented in the following sections. We will use current and projected measurements of \(\kappa_V\) to establish upper bounds on \(\theta\) and, through Eq.~\eqref{eq:key3}, lower bounds on \(m_\eta\), quantitatively probing the viability of the \(\mathrm{SU}(4)/\mathrm{Sp}(4)\) composite Higgs model.
	
	\section{Experimental Constraints on the Vacuum Alignment Angle}
	\label{sec:numerical_analysis}
	
	In the previous sections, the theoretical framework of the minimal \(\mathrm{SU}(4) \to \mathrm{Sp}(4)\) model and the role of vacuum orientation in determining the low-energy Higgs properties were explained in detail. We now turn to converting measured and projected experimental constraints on the Higgs coupling to gauge bosons, \(\kappa_V\), into the space of the vacuum angle \(\theta\) via the fundamental mapping \(\kappa_V = \cos\theta\). We obtain the effective distribution of \(\theta\) and extract one-sided Bayesian credible intervals at \(68\%\) and \(95\%\), together with a frequentist \(\Delta\chi^2\) cross-check. We carry out this procedure for three separate scenarios -- the current ATLAS Run-2 measurement, and the projected precision at the High-Luminosity LHC and at FCC-ee -- so that the same statistical machinery can be compared directly across present and future collider programmes.We then transform each set of limits using the operational mapping defined in Eq.~\ref{eq:meta_theta} into the corresponding lower bound for the mass of the singlet scalar \(\eta\). This relation holds at leading order within the framework of the minimal \(\mathrm{SU}(4) \to \mathrm{Sp}(4)\) model \cite{Cacciapaglia2014}.
	
	We note that Eq.~\eqref{eq:meta_theta} is used here as a leading-order phenomenological relation between two observable quantities -- the singlet mass \(m_\eta\) and the vacuum alignment angle \(\theta\) -- with \(\theta\) treated as a free parameter to be determined from data. While the derivation of this relation in Appendix~\ref{app:mass_relation_detail} proceeds via the minimization condition \ref{eq:min_cond}, the resulting mass ratio \(\frac{m_h^2}{m_\eta^2} = \sin^2\theta\) depends only on the geometric structure of the \(\mathrm{SU}(4)/\mathrm{Sp}(4)\) coset space and not on the specific values of the potential parameters; it therefore retains its validity as \(\theta\) is varied. Higher-order corrections to this relation are expected to be of order \(\theta^2\) and are neglected throughout.
	
	In the numerical method, we first specify the experimental inputs for \(\kappa_V\), each of which enters the analysis as a Gaussian summary \(\mu \pm \sigma\). The first is the comprehensive determination by the ATLAS collaboration based on the Run-2 dataset \cite{ATLAS:2022vkf}. The second is the projected precision on the Higgs vector-boson couplings at the High-Luminosity LHC, as reported jointly by ATLAS and CMS \cite{ATLAS:2025hllhc}. The third is the projected precision at FCC-ee, taken from the baseline running scenario of the FCC Feasibility Study Report \cite{FCC:2025feasibility}. For both projected scenarios, the individual precisions quoted on \(\kappa_W\) and \(\kappa_Z\) are combined in inverse-variance quadrature into a single effective \(\sigma(\kappa_V)\), consistent with the custodial relation \(\kappa_W = \kappa_Z = \cos\theta\) predicted by the model.
	
	For the Bayesian CDF inversion, we map each of these three experimental summaries for \(\kappa_V\) into the angular variable \(\theta\). Under the change of variables \(\kappa_V = \cos\theta\), the Jacobian factor \(|\mathrm{d}\kappa_V/\mathrm{d}\theta| = \sin\theta\) appears in the posterior density. The effective Bayesian posterior density is therefore written as
	
	\begin{equation}\label{fkfhff}
		p(\theta \mid \text{data}) \propto
		\sin\theta \,
		\exp\!\left[-\frac{(\cos\theta - \mu)^2}{2\sigma^2}\right],
		\qquad
		0 \le \theta \le \frac{\pi}{2}.
	\end{equation}
	
	This posterior is used exclusively for the CDF inversion and the resulting Bayesian credible intervals. The frequentist \(\Delta\chi^2\) profile is constructed separately in Sec.~\ref{subsec:methodology} from the corresponding likelihood without Jacobian weighting, and is used only as an independent cross-check.
	
	This expression admits a precise statistical interpretation. Adopting a flat (uniform) prior on \(\kappa_V\), truncated to the physically allowed region \(\kappa_V \leq 1\), and applying the change of variables \(\kappa_V = \cos\theta\) with Jacobian \(\sin\theta\), Eq.~\eqref{fkfhff} is the Bayesian posterior density for \(\theta\) given the data, up to the overall normalization. The factor \(\sin\theta\) therefore plays a dual role: it is simultaneously the Jacobian of the variable transformation and the encoding of the flat prior on \(\kappa_V\).
	
	The extraction of one-sided limits on \(\theta\) is performed based on the inversion of the normalized cumulative distribution function (CDF) to provide a transparent interpretation of credible limits at the \(68\%\) and \(95\%\) levels. Simultaneously, the profile \(\Delta\chi^2(\theta)\) is calculated as a control of the frequentist approach to examine the stability of the results against the choice of statistical procedure \cite{Cowan:2010js}.
	
	After determining the angular limits, the analytical mapping \(m_\eta(\theta) = m_h/\sin\theta\) with \(m_h = 125\ \mathrm{GeV}\) is used to convert the constraints on \(\theta\) into the corresponding lower bound on the singlet mass, separately for each of the three scenarios. Since this transformation is highly nonlinear in the region \(\theta \to 0\), conservative measures — including defining a small numerical cutoff for \(\sin\theta\) and performing grid convergence tests — have been used to protect the results from numerical instabilities. This precaution turns out to matter most for the FCC-ee projection, where the far smaller experimental uncertainty pushes the allowed region for \(\theta\) close to the origin.
	
	The results of this analysis show that, via normalisation of the Bayesian posterior over the physical region \(\kappa_V \leq 1\), upper bounds on the vacuum alignment angle are established for all three scenarios, translating directly into lower bounds on the mass of the singlet \(\eta\) that range from several hundred GeV for the current Run-2 data to the multi-TeV regime once the precision projected for FCC-ee is taken into account.
	
	\subsection{Statistical Framework and Experimental Inputs}
	\label{subsec:methodology}
	
	In this section, we describe the statistical framework used to translate measurements of \(\kappa_V\) into constraints on the vacuum alignment angle \(\theta\). The analysis is based on the leading-order relations derived in Sec.~\ref{sec:theoretical_framework} (Eqs.~\eqref{eq:kappa_theta} and~\eqref{eq:meta_theta}), which connect the observable Higgs coupling modifier to the underlying model parameters \cite{Contino2010, Cacciapaglia2014}. Since the model predicts \(\kappa_W = \kappa_Z = \cos\theta\), we combine separate projections for \(\kappa_W\) and \(\kappa_Z\) in inverse-variance quadrature:
	
	\begin{equation}
		\frac{1}{\sigma(\kappa_V)^2} = \frac{1}{\sigma(\kappa_W)^2} + \frac{1}{\sigma(\kappa_Z)^2},
		\label{eq:kappaV-combination}
	\end{equation}
	
	following standard practice when the full correlation matrix is unavailable.
	
	We consider three experimental scenarios. The first is the current ATLAS Run-2 combined measurement \cite{ATLAS:2022vkf}, reported as \(\kappa_V = 1.035 \pm 0.031\). The second is the projected HL-LHC precision, with \(\sigma(\kappa_W) = \sigma(\kappa_Z) = 1.6\%\) \cite{ATLAS:2025hllhc}, yielding \(\sigma(\kappa_V) \simeq 1.13\%\). The third is the FCC-ee projection, with \(\sigma(\kappa_W) = 0.29\%\) and \(\sigma(\kappa_Z) = 0.10\%\) \cite{FCC:2025feasibility}, giving \(\sigma(\kappa_V) \simeq 0.095\%\). For the projected scenarios, we adopt the Standard Model central value \(\mu = 1\), consistent with the Asimov convention for sensitivity studies.

 As established in Eq.~\eqref{fkfhff}, adopting a flat prior on \(\kappa_V\) truncated to the physical region \(\kappa_V \leq 1\) gives the Bayesian posterior density for \(\theta\), where the factor \(\sin\theta\) plays a dual role as both the Jacobian of the transformation \(\kappa_V=\cos\theta\) and the encoding of the flat prior.  This Bayesian construction is used exclusively for extracting credible intervals via cumulative distribution function (CDF) inversion. As an independent cross-check, we also perform a frequentist profile-likelihood analysis based on the \(\Delta\chi^2\) statistic, constructed directly from the Gaussian likelihood without Jacobian weighting.
	
	A subtle but important point concerns the physical boundary \(\kappa_V \leq 1\). For the Run-2 dataset, the central value \(\mu = 1.035\) lies above this boundary (with approximately $87\%$ of the Gaussian probability mass in the unphysical region $\kappa_V>1$, compared to exactly $50\%$ for the boundary-centered HL-LHC and FCC-ee projections), so  the posterior is significantly affected by the truncation at   \(\kappa_V = 1\). The resulting bounds should therefore be interpreted as conservative upper limits on \(\theta\) rather than as evidence for non-zero vacuum misalignment. For the HL-LHC and FCC-ee projections, by contrast, the Asimov central value \(\mu = 1\) sits exactly on the boundary, so the bounds directly reflect projected experimental sensitivity.
	
	Numerically, we sample the physical range \(\theta \in [0, \pi/2]\) with a uniform grid spacing \(d\theta = 5 \times 10^{-4}\), using a small cutoff \(\theta_{\min} = 10^{-12}\) to avoid boundary instabilities. The posterior is normalized via trapezoidal integration, and credible limits are obtained by inverting the CDF. For the frequentist analysis, one-sided confidence limits are extracted from the intersections of the \(\Delta\chi^2\) profile with \(\Delta\chi^2 = 1\) and \(\Delta\chi^2 = 2.71\), corresponding to \(68\%\) and \(95\%\) confidence levels under Wilks' theorem \cite{Cowan:2010js}. The mass bounds follow from the mapping \(m_\eta = m_h/\sin\theta\), with a numerical floor \(\sin\theta \ge 10^{-12}\) to prevent overflow. Convergence tests confirm that the results are stable against variations in grid spacing and technical cutoffs.
	
	The results of this analysis are summarized in Table~\ref{tab:collider-comparison}, which lists the one-sided Bayesian credible limits and frequentist confidence limits for all three scenarios, together with the corresponding lower bounds on \(m_\eta\).
	
	\begin{table}[h!]
		\centering
		\renewcommand{\arraystretch}{1.4}
		\setlength{\tabcolsep}{6pt}
		\begin{tabular}{lccccc}
			\hline
			\textbf{Scenario} & \(\sigma(\kappa_V)\) & \(\theta_{68\%}\) & \(\theta_{95\%}\) & \(m_\eta^{68\%}\) [GeV] & \(m_\eta^{95\%}\) [GeV] \\
			\hline
			ATLAS Run-2 & 3.10\% & 0.194 & 0.291 & 650 & 440 \\
			HL-LHC      & 1.13\% & 0.150 & 0.211 & 840 & 600 \\
			FCC-ee      & 0.095\% & 0.043 & 0.061 & 2900 & 2100 \\
			\hline
			\textbf{Scenario} & & \(\theta_{\Delta\chi^2=1}\) & \(\theta_{\Delta\chi^2=2.71}\) & \(m_\eta(\Delta\chi^2{=}1)\) & \(m_\eta(\Delta\chi^2{=}2.71)\) \\
			\hline
			ATLAS Run-2 & & 0.153 & 0.232 & 820 & 540 \\
			HL-LHC      & & 0.151 & 0.193 & 830 & 650 \\
			FCC-ee      & & 0.044 & 0.056 & 2900 & 2200 \\
			\hline
		\end{tabular}
		\caption{\footnotesize One-sided Bayesian credible limits (upper block) and frequentist confidence limits (lower block) on \(\theta\), together with the corresponding lower bounds on \(m_\eta\), for the three scenarios considered. The \(\sigma(\kappa_V)\) column lists the combined effective uncertainty entering the analysis.}
		\label{tab:collider-comparison}
	\end{table}
	
	The Bayesian inversion yields \(\theta_{95\%} \simeq 0.291\) for Run-2, corresponding to \(m_\eta \gtrsim 440\ \mathrm{GeV}\), with the frequentist cross-check giving \(\theta \simeq 0.232\) and \(m_\eta \gtrsim 540\ \mathrm{GeV}\). At HL-LHC, the bounds tighten to \(\theta_{95\%} \simeq 0.211\) and \(m_\eta \gtrsim 600\ \mathrm{GeV}\). The FCC-ee projection yields the strongest constraints, with \(\theta_{95\%} \simeq 0.061\) and \(m_\eta \gtrsim 2\ \mathrm{TeV}\). The close agreement between Bayesian and frequentist limits across all scenarios confirms the robustness of these results, which we now explore in detail in the phenomenological discussion that follows.
	
	\section{Phenomenological Implications}
	\label{sec:phenomenology}
	
We now translate the constraints on the vacuum alignment angle obtained in the previous section into concrete phenomenological predictions. The central relations governing this mapping are given in Eqs.~(\ref{eq:kappa_theta}) and (\ref{eq:meta_theta}), which connect the observable Higgs coupling modifier to the singlet scalar mass.
 A small \(\theta\), corresponding to near-perfect alignment, implies a heavy singlet and Higgs couplings close to Standard Model values. The goal of this section is to visualize how the experimental constraints on \(\kappa_V\) translate into bounds on the \(\theta\)--\(m_\eta\) parameter space, compare the reach of current and future colliders, and discuss the implications for the model's testability and naturalness.
	
	\begin{figure}[h!]
		\centering
		\includegraphics[width=0.75\textwidth]{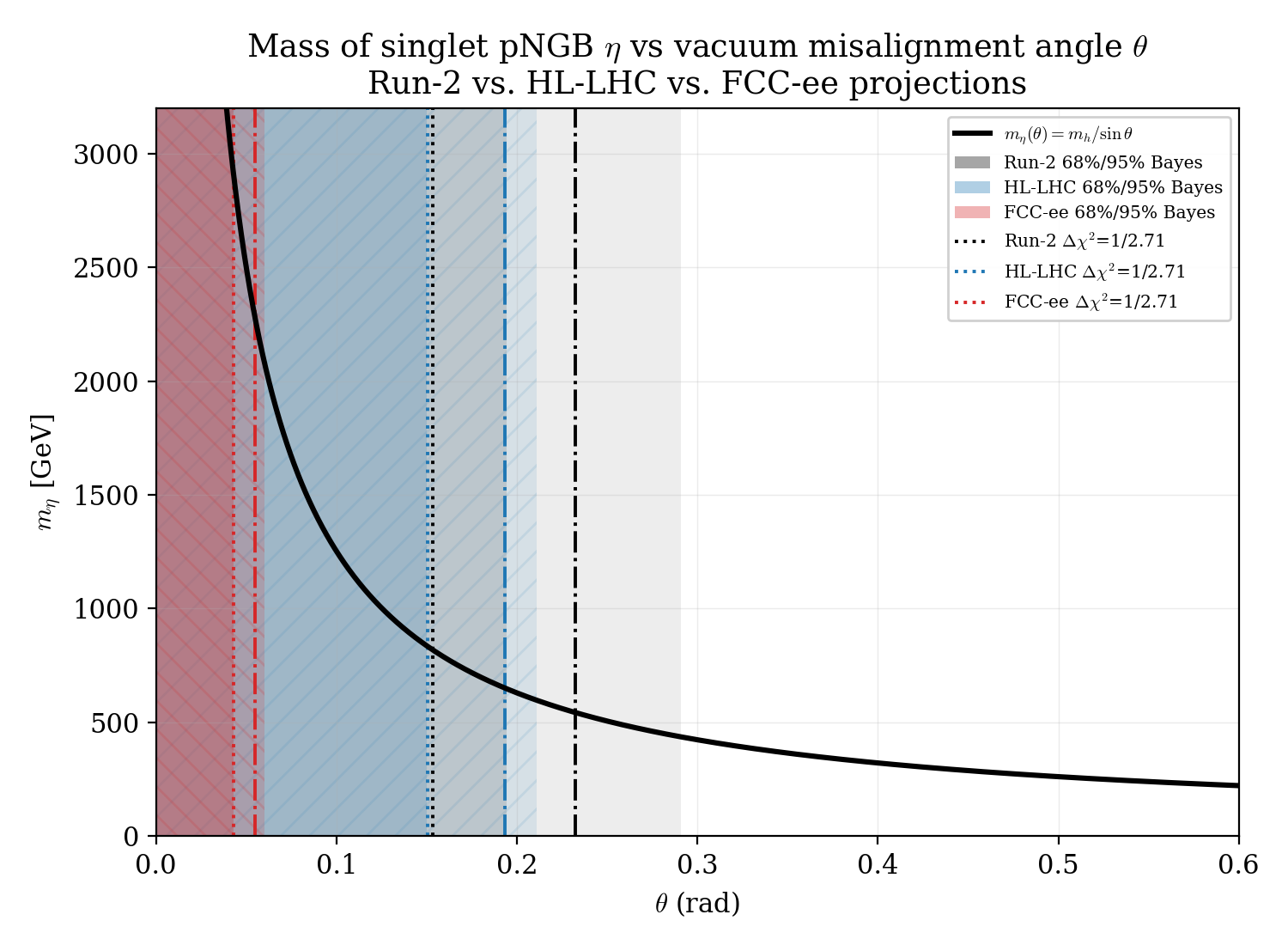}
		\caption{\footnotesize
			\textbf{Mass of the singlet scalar \(\eta\) as a function of the vacuum alignment angle \(\theta\).} The black curve shows the theoretical relation \(m_{\eta}=m_h/\sin\theta\) with \(m_h=125\,\mathrm{GeV}\). The shaded bands indicate the one-sided Bayesian credible limits in \(\theta\) for the three scenarios considered: ATLAS Run-2 (gray), HL-LHC (blue), and FCC-ee (red), with darker shades corresponding to \(68\%\) limits and lighter shades to \(95\%\). The vertical dotted and dash-dotted lines show the corresponding frequentist limits at \(\Delta\chi^2=1\) and \(\Delta\chi^2=2.71\). The progressive narrowing of the allowed region from Run-2 to FCC-ee demonstrates the strengthening of the lower bound on \(m_\eta\) by roughly an order of magnitude.}
		\label{fig:m_eta_vs_theta}
	\end{figure}
	
	Figure~\ref{fig:m_eta_vs_theta} displays the hyperbolic relation \(m_\eta(\theta)=m_h/\sin\theta\). As \(\theta \to 0\), corresponding to perfect alignment, the singlet mass diverges; for larger \(\theta\), \(m_\eta\) approaches \(m_h\). Superimposed on this curve are the Bayesian credible bands derived from the three experimental scenarios. The gray bands for ATLAS Run-2 correspond to \(\theta_{68\%}=0.194\) and \(\theta_{95\%}=0.291\), yielding lower bounds \(m_\eta \gtrsim 650\ \mathrm{GeV}\) and \(m_\eta \gtrsim 440\ \mathrm{GeV}\), respectively. The blue bands for HL-LHC, with \(\theta_{68\%}=0.150\) and \(\theta_{95\%}=0.211\), give \(m_\eta \gtrsim 840\ \mathrm{GeV}\) and \(m_\eta \gtrsim 600\ \mathrm{GeV}\). The red bands for FCC-ee, with \(\theta_{68\%}=0.043\) and \(\theta_{95\%}=0.061\), tighten the bounds dramatically to \(m_\eta \gtrsim 2900\ \mathrm{GeV}\) and \(m_\eta \gtrsim 2100\ \mathrm{GeV}\). The vertical lines show the corresponding frequentist limits from the \(\Delta\chi^2\) profile, which agree closely with the Bayesian bands, particularly for HL-LHC and FCC-ee where the assumed central value \(\mu=1\) lies on the physical boundary. This agreement confirms the robustness of the results across statistical frameworks.
	
	\begin{figure}[h!]
		\centering
		\includegraphics[width=0.75\textwidth]{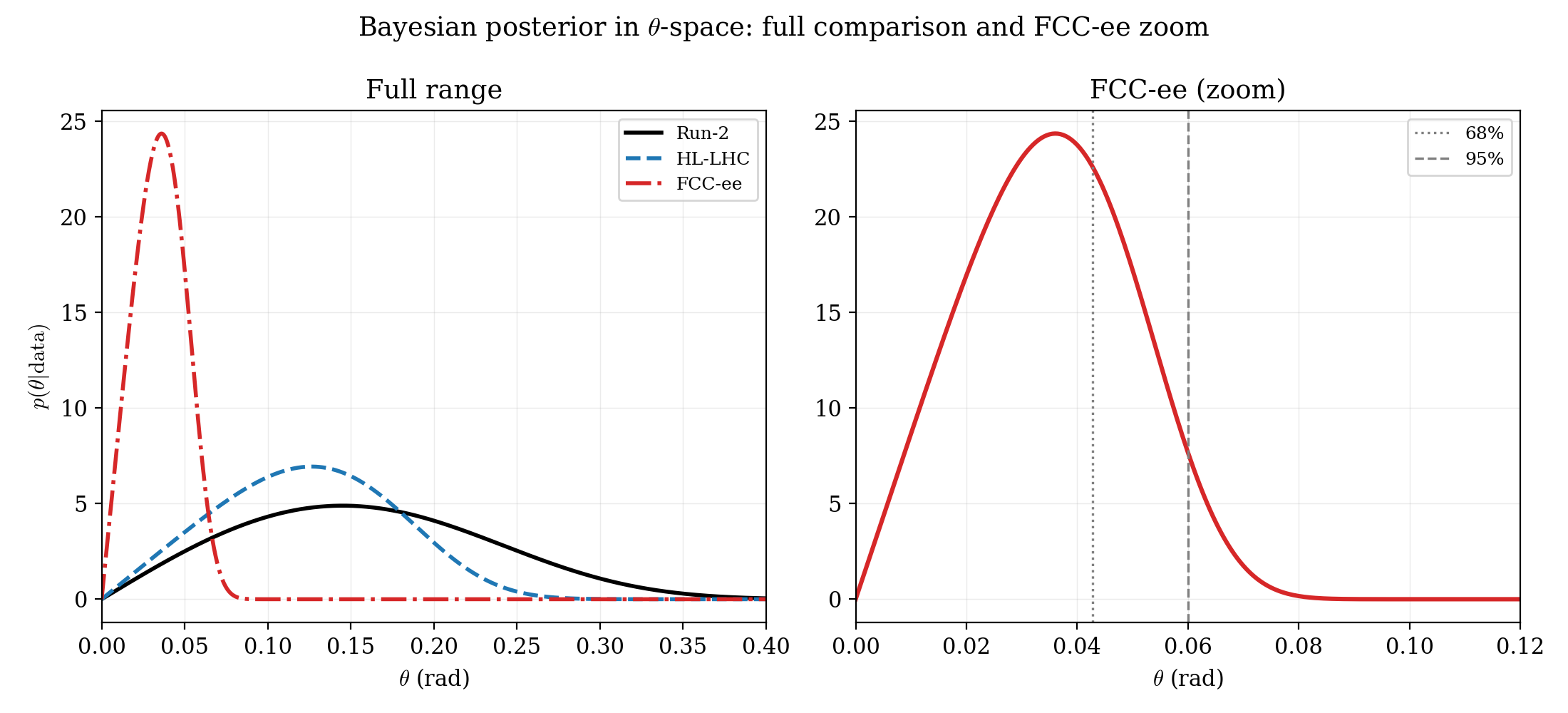}
		\caption{\footnotesize
			\textbf{Normalized Bayesian posterior density for the vacuum alignment angle \(\theta\).} The left panel compares the posterior distributions for ATLAS Run-2 (solid black), HL-LHC (dashed blue), and FCC-ee (dash-dotted red). The right panel provides a magnified view of the FCC-ee posterior, with vertical lines indicating the one-sided credible limits at \(68\%\) (\(\theta_{68\%}=0.043\)) and \(95\%\) (\(\theta_{95\%}=0.061\)). The progressive narrowing of the posterior distributions reflects the increasing sensitivity of Higgs precision measurements to vacuum misalignment.}
		\label{fig:likelihood_profile}
	\end{figure}
	The Bayesian posterior densities shown in Fig.~\ref{fig:likelihood_profile} provide a complementary visualization of how the constraints on \(\theta\) sharpen with improved experimental precision. The posteriors are obtained from Eq.~\eqref{fkfhff}, which includes the Jacobian factor \(\sin\theta\) from the transformation \(\kappa_V=\cos\theta\).
	 This factor forces the posterior to vanish linearly as \(\theta\to0\), producing characteristic non-Gaussian shapes. The broadest distribution corresponds to ATLAS Run-2, where the central value \(\mu=1.035>1\) lies outside the physical region, so the posterior is dominated by the boundary truncation at \(\kappa_V=1\). The HL-LHC projection yields a substantially narrower distribution, while the FCC-ee projection produces an extremely localized posterior concentrated near the origin. The right panel shows the FCC-ee posterior in detail, with the \(68\%\) and \(95\%\) credible limits indicated.
	
	\begin{figure}[h!]
		\centering
		\includegraphics[width=0.75\textwidth]{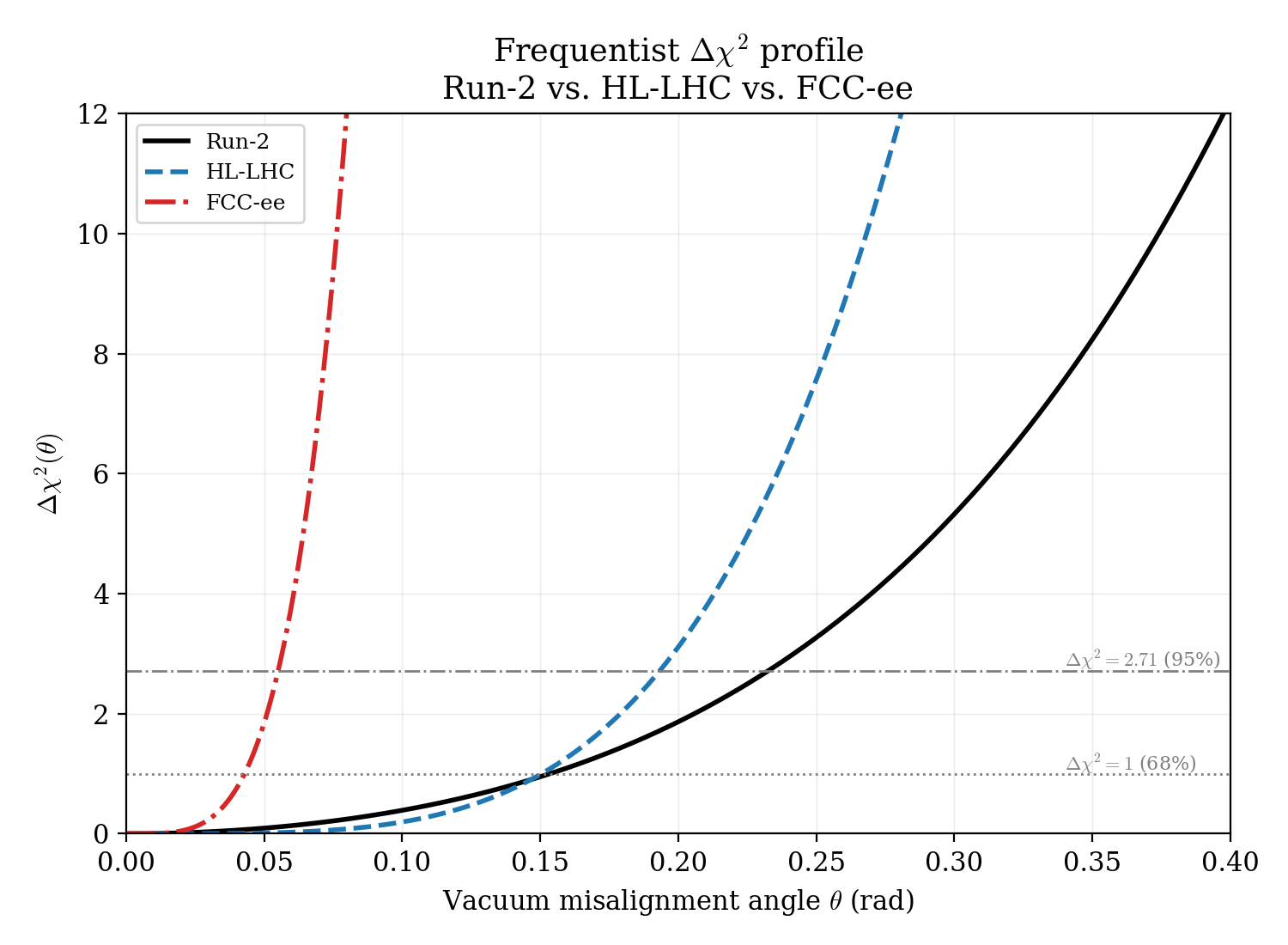}
		\caption{\footnotesize
			\textbf{Frequentist \(\Delta\chi^2\) profile for the vacuum alignment angle \(\theta\).} The three curves correspond to ATLAS Run-2 (solid black), HL-LHC (dashed blue), and FCC-ee (dash-dotted red). The horizontal lines at \(\Delta\chi^2=1\) and \(\Delta\chi^2=2.71\) indicate the approximate one-sided \(68\%\) and \(95\%\) confidence levels. The progressively steeper rise of the profile from Run-2 to HL-LHC to FCC-ee reflects the increasing experimental sensitivity to vacuum misalignment.}
		\label{fig:delta_chi2_profile}
	\end{figure}
	
	The frequentist \(\Delta\chi^2\) profiles in Fig.~\ref{fig:delta_chi2_profile} provide an independent cross-check, constructed directly from the Gaussian likelihood in \(\kappa_V\) space without Jacobian weighting. The intersections with \(\Delta\chi^2=1\) and \(\Delta\chi^2=2.71\) give one-sided confidence limits: for Run-2, \(\theta\simeq0.153\) and \(0.232\) (\(m_\eta \gtrsim 820\) and \(540\ \mathrm{GeV}\)); for HL-LHC, \(\theta\simeq0.151\) and \(0.193\) (\(m_\eta \gtrsim 830\) and \(650\ \mathrm{GeV}\)); and for FCC-ee, \(\theta\simeq0.044\) and \(0.056\) (\(m_\eta \gtrsim 2900\) and \(2200\ \mathrm{GeV}\)). The close agreement with the Bayesian credible intervals across all three scenarios confirms that the conclusions are not artifacts of the statistical framework.
	
	Taken together, these results paint a clear phenomenological picture. The current ATLAS Run-2 measurement already disfavours large vacuum misalignment, pushing the singlet mass above the several-hundred-GeV scale. 	The HL-LHC projection strengthens this bound to approximately \(600\ \mathrm{GeV}\) at the \(95\%\) level, while the FCC-ee projection pushes the lower bound into the multi-TeV regime, excluding \(m_\eta\) below about \(2\ \mathrm{TeV}\). This conclusion is consistent with Ref.~\cite{BelyaevEtAl}, which disfavours light composite singlets using a complementary combination of electroweak precision and LHC data; the present analysis extends that work with an explicit \(\kappa_V\)-to-\(\theta\)-to-\(m_\eta\) mapping specific to the \(\mathrm{SU}(4)/\mathrm{Sp}(4)\) model. This progression has important implications for the testability and naturalness of the model.

	First, the predicted mass range for \(\eta\) lies within reach of current and future collider searches. The discrete symmetry \(\eta \to -\eta\) of the effective Lagrangian forbids tree-level linear couplings of \(\eta\) to Standard Model fields, so decay channels proceed via subleading mechanisms: gauge boson and fermion final states may appear at loop level, while the channel \(\eta \to hh\) becomes accessible if the symmetry is broken by subleading operators or topological terms \cite{Hietanen2014}. These channels, particularly \(\eta \to hh\) for \(m_\eta > 2m_h\), will be primary targets at HL-LHC; at the multi-TeV masses favoured by FCC-ee, direct production would instead require a future high-energy hadron collider, with FCC-ee itself contributing indirectly through the Higgs-coupling precision that drives the bound. Second, the preference for small \(\theta\) bears on naturalness: a small alignment angle corresponds to a large singlet mass and typically requires finer tuning in the effective potential. Consequently, fully natural versions of the model predicting a light \(\eta\) are already disfavoured by current data and would be almost entirely excluded by the precision projected for FCC-ee.
	
	In summary, the phenomenological analysis demonstrates that precision Higgs coupling measurements provide a powerful indirect probe of the \(\mathrm{SU}(4)/\mathrm{Sp}(4)\) composite Higgs model. The progressive tightening of the bounds on \(\theta\) from current LHC data to HL-LHC and FCC-ee translates directly into increasingly stringent lower limits on the singlet mass, ranging from several hundred GeV today to the multi-TeV regime in the future. These results, obtained independently of direct searches, establish the \(\eta\) state as a concrete and well-motivated target for the next generation of Higgs precision programs.
	
	Electroweak precision measurements, which will be sharpened at the FCC-ee project, could also provide competitive
	reach into the misalignment angle $\theta$ \cite{1502.04718}, together with direct searches of $\eta$ at the FCC-ee
	\cite{2104.11064,2211.00961}. However, the $\theta$-dependence in electroweak precision remains a delicate business 
	\cite{1006.0207,1008.1267}.

	\section{Conclusion and Summary}
	\label{sec:conclusion}
	
	In this work, we have investigated the phenomenological implications of the minimal composite Higgs model based on the symmetry-breaking pattern \(\mathrm{SU}(4) \to \mathrm{Sp}(4)\). This model is particularly attractive because it is the simplest realization that yields a Higgs doublet and a singlet scalar \(\eta\) from a fundamental \(\mathrm{SU}(2)\) gauge theory with two Dirac fermions, while preserving custodial symmetry at leading order. Using the leading-order relations \(\kappa_V = \cos\theta\) and \(m_\eta = m_h/\sin\theta\), we translated present and future constraints on the Higgs coupling modifier \(\kappa_V\) into quantitative bounds on the vacuum alignment angle \(\theta\) and the singlet mass \(m_\eta\).
	
	Our numerical framework employed two statistically independent approaches: a Bayesian analysis using the normalized posterior density in \(\theta\)-space, including the Jacobian factor from the transformation \(\kappa_V \to \theta\), and a frequentist profile-likelihood analysis based on the \(\Delta\chi^2\) statistic. The close agreement between the two methods across all three experimental scenarios confirms that the resulting constraints are robust and not artifacts of the statistical framework chosen.
	
	The key numerical results are as follows. Using the current ATLAS Run-2 measurement \(\kappa_V = 1.035 \pm 0.031\), we obtained a Bayesian upper limit \(\theta_{95\%} = 0.291\), corresponding to \(m_\eta \gtrsim 440\ \mathrm{GeV}\), with the frequentist cross-check giving \(\theta \simeq 0.232\) and \(m_\eta \gtrsim 540\ \mathrm{GeV}\). For the projected HL-LHC precision, the Bayesian constraint improves to \(\theta_{95\%} \simeq 0.211\), pushing \(m_\eta\) above approximately \(600\ \mathrm{GeV}\). The improvement at FCC-ee is far more dramatic: the Bayesian limit tightens to \(\theta_{95\%} \simeq 0.061\), implying a lower bound \(m_\eta \gtrsim 2100\ \mathrm{GeV}\), with the frequentist analysis giving consistent results. The Run-2 bounds, however, should be interpreted with caution: since the measured central value lies above the physical boundary \(\kappa_V \leq 1\), these limits are primarily boundary-driven and conservative rather than evidence for non-zero vacuum misalignment.
	
	The main conclusion of this study is clear and sharpens considerably when future data are considered. Increasingly precise Higgs coupling measurements translate directly into increasingly stringent limits on vacuum misalignment and progressively push the singlet scalar mass into the heavy-mass regime. While current data constrain \(\eta\) to lie above the several-hundred-GeV scale, the precision expected at FCC-ee would probe the model at the multi-TeV level through indirect measurements alone —a qualitatively different regime that would essentially close off light, strongly misaligned realizations of the model.
	
	These results have important implications for the naturalness of the composite Higgs framework. The growing preference for small \(\theta\), driven by the improving precision of \(\kappa_V\), corresponds to a large singlet mass \(m_\eta = m_h/\sin\theta\). Achieving such a small alignment angle typically requires increasingly fine-tuned cancellations in the effective potential. Consequently, fully natural versions of the model that predict a light \(\eta\) are already disfavoured by current data and would be almost entirely excluded by the precision projected for FCC-ee. This tension between precision Higgs data and naturalness highlights a persistent challenge for the composite Higgs program — one that future colliders will only sharpen further.
	
	The minimal \(\mathrm{SU}(4)/\mathrm{Sp}(4)\) composite Higgs framework thus remains both predictive and testable. Future progress will come from the combined power of precision Higgs measurements, direct searches for heavy scalar resonances at HL-LHC and future hadron colliders, and improved theoretical control of the underlying strongly coupled dynamics through lattice simulations and effective field theory. Together, these developments will provide increasingly decisive tests of the composite origin of electroweak symmetry breaking and will determine whether the \(\eta\) singlet — or its absence — reveals the nature of new physics beyond the Standard Model.
	
	\section*{Acknowledgements}
	
	We would like to thank Dr. Moslem Ahmadvand for valuable discussions. 
	\appendix
	
\section{Complete Derivation of the $SU(4) \to Sp(4)$ Symmetry Breaking from the Lagrangian}
\label{app:lagrangian_rewrite}

In this appendix we outline the derivation of the $\mathrm{SU}(4)\to\mathrm{Sp}(4)$ breaking pattern from the Lagrangian of the $\mathrm{SU}(2)$ gauge theory with two Dirac flavors, adopting the notation of Sec.~\ref{sec:theoretical_framework}. The four left-handed Weyl spinors are collected into the column vector

\begin{equation}
	Q_I^{\,i\alpha} = \begin{pmatrix} U_L^{\,i\alpha} \\ D_L^{\,i\alpha} \\ \widetilde U_L^{\,i\alpha} \\ \widetilde D_L^{\,i\alpha} \end{pmatrix}, \qquad I=1,\dots,4,
\end{equation}

where the charge-conjugated fields $\widetilde U_L$ and $\widetilde D_L$ are defined via $\widetilde U_L^{\,i\alpha} = \varepsilon^{ij} C^{\alpha\beta} (\overline{U_R})^T_{j\beta}$, and similarly for $D$. The mass term for the two Dirac fermions can be rewritten as

\begin{equation}
	m(\bar U U + \bar D D) = \frac{m}{2} Q^T (\varepsilon \otimes C) E Q + \frac{m}{2} \big(Q^T (\varepsilon \otimes C) E Q\big)^\dagger,
\end{equation}

where $\varepsilon$ is the antisymmetric color tensor, $C$ is the charge conjugation matrix, and

\begin{equation}
	E = \begin{pmatrix}
		0 & 0 & 1 & 0 \\
		0 & 0 & 0 & 1 \\
		-1 & 0 & 0 & 0 \\
		0 & -1 & 0 & 0
	\end{pmatrix}.
\end{equation}

This rewriting makes the $\mathrm{SU}(4)$ flavor symmetry manifest in the massless limit. For an infinitesimal transformation $Q \mapsto (\mathbb{I} + i\alpha_n T_n)Q$, the mass term transforms as

\begin{equation}
	\delta\big(Q^T(\varepsilon\otimes C)E Q\big) = i\alpha_n\, Q^T(\varepsilon\otimes C)(E T_n + T_n^T E)Q + \mathcal{O}(\alpha^2).
\end{equation}

Invariance requires $E T + T^T E = 0$, which is the defining condition for the Lie algebra $\mathfrak{sp}(4)$.

 Writing the $4\times4$ generator $T$ in $2\times2$ block form, 
\begin{equation}
	T=\begin{pmatrix}p&q\\ r&s\end{pmatrix},
\end{equation}
 and imposing $T^\dagger=-T$, $\mathrm{Tr}\,T=0$, together with $ET+T^TE=0$, one finds $s=-p^T$ and $q^T=q$, $r^T=r$. Counting parameters, $p$ contributes 4 real parameters and each of the symmetric matrices $q,r$ contributes 3, giving $4+3+3=10$ generators in total -- matching the dimension of $\mathfrak{sp}(4)$.  By the Nambu--Goldstone theorem, the remaining five generators of $\mathfrak{su}(4)$ that do not satisfy this condition are broken, yielding the five pseudo-Nambu--Goldstone bosons discussed in the main text.
Thus, the symmetry is broken as $\mathrm{SU}(4) \to \mathrm{Sp}(4)$, and the resulting pNGBs decompose under the electroweak subgroup into a Higgs doublet and a neutral singlet $\eta$, as described in Sec.~\ref{sec:theoretical_framework}.
	
	\section{Derivation of the Mass Relation \texorpdfstring{$m_\eta = m_h/\sin\theta$}{m\_eta = m\_h/sin theta}}
	\label{app:mass_relation_detail}
	
	This appendix presents the derivation of the mass relation \(m_\eta = m_h/\sin\theta\), which arises from the effective potential generated by the explicit breaking of the global \(\mathrm{SU}(4)\) symmetry. The dominant sources of this breaking are the electroweak gauge interactions and the top-quark Yukawa coupling.
	
	\subsection*{Gauge Contribution}
	
	The gauge contribution to the effective potential, generated by loops of electroweak gauge bosons, takes the form
	
	\begin{equation}
		V_{\rm gauge} = -C_g f^4 \left( g^2 \sum_{i=1}^{3} \mathrm{Tr}\!\left(S^i \cdot \Sigma \cdot (S^i \cdot \Sigma)^*\right) + g'^2 \mathrm{Tr}\!\left(S^6 \cdot \Sigma \cdot (S^6 \cdot \Sigma)^*\right) \right),
	\end{equation}
	
	where \(C_g > 0\) is a low-energy constant. Expanding to second order in the fields gives
	
	\begin{equation}
		V_{\rm gauge} \sim C_g \frac{3g^2 + g'^2}{2} \left( -f^4 \cos^2\theta + \frac{f^3}{\sqrt{2}} \cos\theta \sin\theta\, h + \frac{f^2}{8}\big(\cos 2\theta\, h^2 - \sin^2\theta\, \eta^2\big) + \cdots \right).
	\end{equation}
	
	Since \(C_g > 0\), this contribution alone is minimized at \(\theta = 0\), driving the vacuum toward the electroweak-preserving configuration.
	
	\subsection*{Top-Quark Contribution}
	
	The dominant contribution arises from top-quark loops. The top-quark mass is generated through a four-fermion operator, leading to a top-loop contribution to the potential:
	
	\begin{equation}
		V_{\rm top} = -C_t y_t'^2 f^4 \sum_{\alpha=1}^{2} \left[\mathrm{Tr}(P^\alpha \Sigma)\right]^2,
	\end{equation}
	
	where \(C_t > 0\) and \(P^\alpha\) are projectors onto the \(\mathrm{SU}(2)_L\)-doublet components. Expanding to second order:
	
	\begin{equation}
		V_{\rm top} \sim -C_t y_t'^2 \left( f^4 \sin^2\theta + \frac{f^3}{\sqrt{2}} \cos\theta \sin\theta\, h + \frac{f^2}{8}\big(\cos 2\theta\, h^2 - \sin^2\theta\, \eta^2\big) + \cdots \right).
	\end{equation}
	
	This contribution is minimized at \(\theta = \pi/2\), favoring the Technicolor-like vacuum.
	
	\subsection*{Explicit Breaking}
	
	A third contribution comes from an explicit mass term for the techni-fermions, aligned with the fixed vacuum direction \(\Sigma_B\):
	
	\begin{equation}
		V_m = C_m f^4 \, \mathrm{Tr}(\Sigma_B \cdot \Sigma) \sim C_m \left( -4f^4 \cos\theta + \sqrt{2} f^3 \sin\theta\, h + \frac{1}{4} f^2 \cos\theta\, (h^2 + \eta^2) + \cdots \right),
	\end{equation}
	
	where \(C_m\) may take either sign, providing the freedom to shift the vacuum toward small \(\theta\).
	
	\subsection*{Combined Potential and Mass Spectrum}
	
	Summing the three contributions and defining
	
	\begin{equation}
		X_t \equiv y_t'^2 C_t - \frac{3g^2 + g'^2}{2} C_g, \qquad X_m \equiv C_m,
	\end{equation}
	
	the potential simplifies to
	
	\begin{equation}
		V(\theta) = X_t \cos^2\theta - 4X_m \cos\theta + \text{const.}
	\end{equation}
	
	Minimizing with respect to \(\theta\) gives
	
	\begin{equation}
		\cos\theta_{\min} = \frac{2X_m}{X_t},
	\end{equation}
	
	which is physical only if \(X_t > 2|X_m|\). The second derivatives at the minimum yield the mass eigenvalues:
	
	\begin{equation}
		m_h^2 = \frac{f^2}{4} X_t \sin^2\theta, \qquad m_\eta^2 = \frac{f^2}{4} X_t.
	\end{equation}
	
	Both masses share the same prefactor \(f^2 X_t/4\); the only difference is the geometric factor \(\sin^2\theta\) multiplying the Higgs mass. Dividing the two expressions gives
	
	\begin{equation}
		\frac{m_h^2}{m_\eta^2} = \sin^2\theta,
	\end{equation}
	
	and hence
	
	\begin{equation}
		m_\eta = \frac{m_h}{\sin\theta}.
	\end{equation}
	
	This relation is a direct consequence of the group-theoretic structure of the \(\mathrm{SU}(4)/\mathrm{Sp}(4)\) coset: the Higgs direction couples to the explicit-breaking sources in a way that tracks the alignment angle, while the singlet direction is orthogonal and retains its mass even as \(\theta \to 0\). The relation holds at leading order in the effective theory; higher-order corrections are expected to be of order \(\theta^2\) and are neglected in this work.

\end{document}